\documentclass[oldversion]{aa}
\usepackage{epsfig,psfig}
\usepackage{txfonts}
\usepackage{color}
\usepackage{graphicx}

\newcommand{\msun}{\mbox{$M_{\odot}$}}
\newcommand{\Msun}{\mbox{$M_{\odot}$}}
\newcommand{\lsun}{\mbox{$L_{\odot}$}}

\newcommand{\zsun}{\mbox{$Z_{\odot}$}}
\newcommand{\teff}{\mbox{$T_{\rm eff}$}}

\newcommand{\vinf}{\mbox{$\varv_{\infty}$}}
\newcommand{\vesc}{\mbox{$\varv_{\rm esc}$}}

\newcommand{\mdot}{\mbox{$\dot{M}$}}

\newcommand{\msunyr}{\mbox{$M_{\odot} {\rm yr}^{-1}$}}

\def\aap{A\&A}                
\def\apj{ApJ}                 
\def\apjl{ApJ}                
\def\apss{Ap\&SS}             
\def\araa{ARA\&A}             
\def\mnras{MNRAS}             
\def\nat{Nature}              
\def\pasj{PASJ}               
\def\pasp{PASP}               
\newcommand{\realcleardoublepage}{\clearpage
  \ifodd \arabic{page}\else \thispagestyle{empty}\mbox{}\newpage \fi }

\newcommand{\be}{\begin{equation}}
\newcommand{\ee}{\end{equation}}

\newcommand{\mstar}{\mbox{$M_{\star}$}}

\newcommand{\kmsec}{\mbox{km\,s$^{-1}$}}

\begin{document}

\title{Very massive stars: a metallicity-dependent upper-mass limit, slow winds, and 
the self-enrichment of globular clusters}

\author{Jorick S. Vink\inst{1}}
\offprints{Jorick S. Vink, jsv@arm.ac.uk}

\institute{Armagh Observatory and Planetarium, College Hill, Armagh, BT61 9DG, Northern Ireland}

\titlerunning{Metallicity-dependent winds from very massive stars}
\authorrunning{Jorick S. Vink}

\abstract{One of the key questions in Astrophysics concerns the issue of whether there exists
an upper-mass limit to stars, and if so, what physical mechanism sets this limit? The answer to this question 
might also determine if the upper-mass limit is metallicity ($Z$) dependent. 
We argue that mass loss by radiation-driven winds mediated by line opacity is one 
of the prime candidates setting the upper-mass limit. 
We present mass-loss predictions ($\mdot_{\rm wind}$) from Monte Carlo radiative 
transfer models for relatively cool (\teff\ $=$ 15kK) inflated 
very massive stars (VMS) with large Eddington $\Gamma$ factors 
in the mass range $10^2-10^3$ \msun\ as a function of metallicity down to 1/100 $Z/\zsun$. We 
employed a hydrodynamic version of our Monte Carlo method, allowing us to predict the rate 
of mass loss ($\mdot_{\rm wind}$) and the terminal wind velocity (\vinf) simultaneously. 
Interestingly, we find wind terminal velocities (\vinf) that are low (100-500 km/s) 
over a wide $Z$-range, and we propose that the slow winds from VMS are an important
source of self-enrichment in globular clusters.
We also find mass-loss rates ($\mdot_{\rm wind}$), exceeding the typical mass-accretion 
rate ($\mdot_{\rm accr}$) of $10^{-3}$ \msunyr\ during massive-star formation. 
We have expressed our mass-loss predictions as a function of mass and $Z$, 
finding $\log$ \mdot\ $=$ $-$9.13 $+$ 2.1 $\log(M/\Msun)$ $+$ 0.74 $\log(Z/\zsun)~(M_{\odot}/{\rm yr})$.
Even if stellar winds do not directly halt \& reverse mass accretion during star formation, 
if the most massive stars form by stellar mergers, stellar wind mass loss may dominate over the rate at which 
stellar growth takes place. We therefore argue that the upper-mass limit is effectively $Z$-dependent due to
the nature of radiation-driven winds. This has dramatic consequences for the most luminous supernovae, gamma-ray bursts, 
and other black hole formation scenarios at different Cosmic epochs.}


\maketitle


\section{Introduction}
\label{s_intro}

We present mass loss-predictions for very massive stars (VMS) in the $10^2-10^3$ \msun\ range, which
may also provide useful insights for winds from supermassive stars (SMS) above $10^3$ \msun\ range that 
may exist in the dense centres of globular clusters (e.g. Portegies-Zwart et al. 2004). 
SMS have been suggested to be responsible for the observed anti-correlations 
in stellar abundances (between Na and O; Mg and Al) of low-mass stars in 
globular clusters (Denissenkov \& Hartwick 2014; Gieles et al. 2018). 
Here we propose our predicted slow winds from VMS as a source of internal pollution of Globular Clusters 
as an alternative. 

Until 2010, most astronomers thought the stellar initial mass function (IMF) had an upper limit in the
range 100-150\msun\ (Weidner \& Kroupa 2004; Figer 2005) and for this reason early investigations of potential polluters
of Globular clusters by Prantzos \& Charbonnel (2006) only included stars up to 100\msun\ in their
analysis of possible pollution by the winds from massive stars as an alternative to massive asymptotic giant branch (AGB) stars. We have recently discovered 
in the context of the VLT-Flames Tarantula Survey (VFTS; Evans et al. 2011; Vink et al. 2017) 
that the number of massive stars above 30\msun\ is significantly larger than expected from a Salpeter 
IMF (Schneider et al. 2018). But there is more: within the same VFTS survey we found evidence 
for an upturn in the mass-loss rates of VMS above $\sim$100\msun\ (Bestenlehner et al. 2014), as
predicted (Vink et al. 2011). These stars are identified as WNh stars 
(Crowther et al. 2010; Gr\"afener
et al. 2011; Martins 2015) with a nitrogen-enhanced Wolf-Rayet appearance, but still with hydrogen (H)
present, as expected for H-burning main sequence stars. Thus, in contrast to the hypothetical SMS, we can be 
sure that VMS exist in Nature.

Regarding globular clusters, one of the main problems with most of the proposed (AGB or massive star) 
self-enrichment sources 
from nucleo-synthesis is the so-called mass budget problem (Bastian \& Lardo 2018). However, given the 
discovered excess of very massive stars, 
enhancing the kinetic wind energy and momentum by at least a factor of five (Schneider et al. 2018), as well as the 
enhanced stellar wind strength from these very massive stars, VMS
may be both plentiful and individually powerful enough to provide the required amount of 
mass loss to overcome the mass-budget problem, naturally becoming the main contender for being the  
source of the chemical pollution of globular clusters.
So far however, the most fundamental opposition to the winds of (very) massive stars has been the property of 
their fast (2000-3000 km/s) outflow speeds, significantly larger than the potential wells 
of globular clusters would allow (e.g. Decressin et al. 2007; de Mink et al. 2009; Gieles et al. 2018). 
However, here we predict slow winds from VMS
for cool ($\simeq$ 15 kK) and inflated VMS, allowing the winds from VMS to be the most natural source for the self-enrichment
of globular clusters. 

One of the key questions in astrophysics concerns the question of whether there exists
an upper mass limit to stars, and what physical mechanism may set such an upper limit.
Prior to the inclusion of the OPAL opacities in stellar structure models, even at solar $Z$ it was  
possible to construct stellar models for stars with $10^6$\Msun\ or more (e.g. Kato 1986), but 
this limit has dropped to $10^3$\Msun\ over the last decade (Ishii et al. 1999, 
Yungelson et al. 2008). This still does not mean that stars of $10^3$\Msun\ will actually form in nature at solar $Z$,
as the actual upper limit may be dependent on various feedback effects (Krumholz 2015; Tanaka et al. 2017) of which
stellar winds is one of them.

For many decades it appeared almost impossible to form massive stars without the need to resort to stellar mergers, as 
radiation pressure on dust grains could reverse the infalling material -- limiting the stellar mass to a maximum value as low as 
$\sim$10\msun\ (e.g. Larson \& Starrfield 1971; Wolfire \& Cassinelli 1987). However, in these early models the accretion was assumed to be 
spherically symmetric -- an unlikely scenario in nature. Indeed, more recent multi-D simulations 
(Yorke \& Sonnhalter 2002; Krumholz et al. 2009; Kuiper et al. 2010; Rosen et al. 2016) 
indicate there is no fundamental problem in growing (very) massive stars via equatorial 
accretion disks. The only limiting factor seems to be the amount of material 
initially available for the hydrodynamical simulations (Krumholz 2015). 

We note that the star-formation simulations do not actually resolve the innermost grid-point, and 
it is stellar physics that ultimately determines the fate of the object. 
For example, the above-mentioned 
numerical simulations only include dust opacity, but do not include the line opacity of atomic gas -- known
to be the dominant opacity source in mass-loss computations for massive stars, since the early 1970s 
(Lucy \& Solomon 1970, Castor et al. 1975 (CAK), Pauldrach et al. 1986, Vink et al. 2000, Krticka et al. 2016; M\"uller \& Vink 2008).

Tanaka et al. (2017) recently showed that given typical mass-{\it accretion} rates of order $10^{-3}$ \msunyr, feedback
mechanisms need to be of the same order of magnitude as this high mass-accretion rate to significantly affect the maximum 
stellar mass. They considered mass-loss rates from stellar winds for hot stars ($\simeq$ 50 kK), and estimated these 
to be too low.
However, VMS are expected to be `bloated' (Hosokawa \& Omukai 2009)
before they reach the zero-age main sequence (ZAMS), for instance due to envelope inflation due to the Fe opacity
bump (Ishii et al. 1999; Gr\"afener et al. 2012). We should therefore analyse mass-loss rates at 
lower effective temperatures (\teff\ $=$ 15 kK) below the so-called wind
bi-stability jump (Pauldrach \& Puls 1990; Vink et al. 1999) where, as we show in this paper, the rates 
can reach the same order of magnitude as the typically assumed mass-accretion rate of $10^{-3}$ \msunyr. 
This suggests that metallicity-dependent stellar winds ultimately cause the upper-mass 
limit to be $Z$ dependent. 

Interestingly, the most massive stars
may possibly form by stellar collisions between lower mass stars in dense clusters (e.g. Portegies Zwart et al. 2010; Banerjee et al. 2012).
This scenario may possibly lead to the formation of 1000 \msun\ VMS if mass loss during stellar evolution
were not important. However, Belkus et al. (2007); Yungelson et al. (2008); Pauldrach et al. (2012); Yusof et al. (2013) and 
K\"ohler et al. (2015) showed
that the `effective' upper mass limit is expected to be far lower than 1000 \msun\ due to stellar wind mass loss
at solar metallicity.  

Here we investigate if the mass-loss rates of VMS in the $10^2-10^3$ \msun\ range 
are Z-dependent, arguing for a $Z$-dependent effective upper mass limit.
VMS are observed to have masses
up to 200-300 $\msun$, which appears to be the currently known empirical
upper-mass limit (Crowther et al. 2010; Bestenlehner et al. 2014; Martins 2015).
Such stars are no longer identified as O-type stars, but WNh stars (Wolf-Rayet stars with Nitrogen and Hydrogen), which
simply form the extension of the main sequence to the highest known empirical stellar masses (Vink et al. 2015), 
forming a 
natural sequence of O stars, Of/WN transition stars, and the most massive WNh emission line stars due to higher \& higher mass loss. 
Although there is no {\it need} to invoke that these WNh stars are stellar mergers above the canonical
Figer et al. (2005) 150 \msun\ limit, it is of course well possible that mergers (either as dynamical collisions, or
via binary evolution) could contribute to making the most massive stars.

VMS are key for correctly predicting the ionizing radiation of these hot stars, with 
major consequences for interpreting He {\sc ii} line emission at intermediate and high red-shifts as 
Population {\sc iii} stars (Cassata et al. 2013; Sobral et al. 2015), or very massive stars
(VMS) at low metallicity (Gr\"afener \& Vink 2015). 

This paper is organized as follows. 
In Sect.~\ref{s_model}, we briefly describe the Monte Carlo modelling, and  
the parameter space considered for this study (Sect.\,\ref{s_param}).
The mass-loss predictions (Sect.~\ref{s_res}) for cool VMS are followed by discussions on the
relevance for massive star formation feedback and the upper-mass limit (Sect.\,\ref{s_upper}) and 
the self-enrichment of globular clusters (Sect.\,\ref{s_self}), 
before ending with a summary in Sect.~\ref{s_sum}.


\section{Physical assumptions and Monte Carlo modelling}
\label{s_model}

In this paper we predict mass loss due to 
stationary stellar winds on the basis of the Castor, Abbott, \& Klein (1975)
radiation-driven wind model, with modifications to allow 
for multiple-line scattering. 
The details of our approach -- inspired by 
Abbott \& Lucy (1985) --
are given in Vink et al. (1999 and references therein). 
As in our previous computations of mass-loss rates for hotter 
VMS (Vink et al. 2011) we 
employ our dynamical approach (M\"uller \& Vink 2008, 2014; 
Muijres et al. 2012), in which we predict $\mdot$, $\vinf$ and the wind structure 
parameter $\beta$ simultaneously.

The underlying model atmosphere code {\sc isa-wind} (de Koter et al. 1993) 
computes the chemical elements H, He, C, N, O, S, and Si explicitly in non-LTE. 
The iron-group elements are treated in the simplified approach of Schmutz (1991). 
Tests were performed in which we treated 
Fe explicitly in non-LTE, but as this showed only very small differences
with respect to the simplified models, we kept
treating Fe in the approximate way.           
The Kurucz \& Bell (1995) line list includes million of lines, of which we selected 
the strongest $10^5$ transitions of the first 30 elements in the periodic table. 

We have assumed the winds to be spherically symmetric and homogeneous. 
For hot massive stars in the canonical mass range, wind clumping is known to have a profound effect (Hillier 1991), leading to a downwards adjustment 
of empirical mass-loss rates, by a factor 
of $\simeq$3 (Moffat \& Robert 1994; Puls et al. 2008; 
Hamann et al. 2008; Ram\'irez-Agudelo et al. 2017).

Wind clumping may also have a theoretical effect on the radiative driving.
Furthermore, given that VMS are relatively close to the Eddington limit there 
could be additional
physics resulting in the development of porous structures (van Marle et al. 2008; Gr\"afener et al. 2012;
Jiang et al. 2015).
The issue of porosity on mass-loss rate predictions 
for O-type stars was studied by Muijres et al. (2011), where it was found that whilst 
the impact on \mdot\ can be large for certain clumping prescriptions, the 
overall conclusion for moderate 
clumping and porosity is that it does not change the mass-loss predictions substantially.
In the present set of computations we do therefore not
account for wind clumping. 

Proximity to the Eddington limit may also lead to eruptive mass loss in luminous blue variable (LBV) 
Eta Car type giant 
eruptions (e.g. Shaviv et al. 2000; Smith \& Owocki 2006; Owocki 2015). 
This may lead to additional mass loss, which may be less dependent on metallicity (although due to 
the Fe opacity bump the metallicity may still play a role).
We note however that the model-independent transition mass-loss rate of Vink \& Gr\"afener (2012), which
allowed a calibration of stationary mass-loss rates, showed that giant eruptions are not required to offset
the lower mass-loss rates from clumped stationary winds. 
However, if LBV-type giant eruptions {\it do} occur sufficiently frequently (and vigorously), they might 
provide an additional avenue for mass loss, on top of what is predicted in this paper.

\section{Parameter space}
\label{s_param}

We managed to converge models up to 900\msun\ within the M\"uller \& Vink (2008) dynamical framework, to 
estimate the mass-loss rates and wind velocities of VMS.
For a range of masses, stellar luminosities were derived from the mass-luminosity 
relation of Gr\"afener et al. (2011), forming the basis of our grid. We added a few additional models where 
the stellar mass was varied with respect to the standard luminosity value, thus changing the 
$M$-$L$ ratio, and thus the Eddington $\Gamma$ value (see the bottom of Table\,1). We note that we refer
to the Eddington factor for electron scattering only, but for a more extensive discussion on the total 
opacity see Vink et al. (2011).

The effective temperature sets the ionization stratification in the atmosphere and 
determines which lines are most active in driving the wind. As a result, \teff\
affects the predicted mass-loss rate. Here, we fix \teff\ to 15\,000\,K. 
The reason for these cool temperatures is envelope inflation, which is thought to occur 
both during pre-main sequence (PMS) and post-main sequence evolution (Tanaka et al. 2017; Gr\"afener et al. 2012).
We do not express the mass-loss rates 
as a function of \teff, as radii are highly uncertain due to the likely 
clumped nature of inflated stellar envelopes (Gr\"afener et al. 2012).

Our grid has been constructed to predict the mass-loss behaviour as a function of $M$ 
(or $\Gamma$ via the $M$-$L$ relation) and $Z$, which are scaled to the solar values 
(Anders \& Grevesse 1989), and investigated down to 1/100 $Z/\zsun$, as relevant 
for VMS in the present-day universe, as well as for globular cluster metallicities (see Sect.\,\ref{s_self}).

\section{Results}
\label{s_res}

\begin{table*}
\centering
\begin{tabular}{llll|l|ccc}
\hline
\hline\\[-6pt]
\mstar  & $\log L$ & $\Gamma$ & $Z/\zsun$ & \vesc    & \vinf   & $\log \mdot$ & \(\beta\) \\[2pt]

[\msun] & [\lsun] &           &           & [\kmsec] & [\kmsec] & [\msunyr]   &         \\[3pt]
\hline\\[-7pt]
100    &  6.11  & 0.312     & $1$       & 478 &    635 & $-$4.97 &  0.87\\
       &        &           &           & &   1548 & $-$5.72 &  1.0  \\
\hline
       &        &           & $1/3$     & &    584 & $-$5.32 & 0.78 \\
       &        & 0.317     & $1/10$    & &    554 & $-$5.74 & 0.63 \\
       &        &           & $1/33$    & &    441 & $-$6.02 & 0.63 \\
       &        & 0.318     & $1/100$   & &    655 & $-$6.62 & 0.80 \\
150    & 6.38   &           & $1/3$     & 500 &    657 & $-$4.97 & 0.86 \\
       &        & 0.396     & $1/10$    & &    479 & $-$5.28 & 0.69 \\
       &        &           & $1/33$    & &    439 & $-$5.62 & 0.61 \\
       &        & 0.397     & $1/100$   & &   584 & $-$6.05 & 0.77 \\
200    & 6.56   & 0.455     & $1/10$    & 519 &   527 & $-$5.08 & 0.67 \\
       &        & 0.456     & $1/100$   & &   613 & $-$5.81 & 0.72 \\
250    & 6.70   & 0.504     & $1/10$    & 536 &   462 & $-$4.79 & 0.70\\
       &        &           & $1/33$    & &   415 & $-$5.16 & 0.63 \\
       &        & 0.504     & $1/100$   & &   571 & $-$5.57 & 0.74 \\
300    & 6.81   & 0.541     & $1/10$    & 550 &   490 & $-$4.66 & 0.78 \\
       &        &           & $1/33$    & &   454 & $-$5.05 & 0.65 \\
350    & 6.91   & 0.573     & $1/10$    & 564 &   482 & $-$4.50 & 0.73 \\
       &        &           & $1/33$    & &   419 & $-$4.87 & 0.64 \\
       &        & 0.574     & $1/100$   & &   565 & $-$5.26 & 0.74 \\
400    & 6.99   & 0.601     & $1/10$    & 576 &   426 & $-$4.36 & 0.71\\
       &        &           & $1/33$    & &   484 & $-$4.82 & 0.65 \\
       &        & 0.603     & $1/100$   & &   537 & $-$5.13 & 0.71 \\
500    & 7.11   & 0.645     & $1/10$    & 599 &   339 & $-$4.13 & 0.86 \\
550    & 7.17   &           & $3$       & 609 &   507 & $-$3.03 & 0.65 \\
       &        & 0.652     & $1$       & &   111 & $-$3.21 & 0.63 \\
       &        &           & $1/3$     & &   310 & $-$3.72 & 0.67 \\
       &        & 0.664     & $1/10$    & &   344 & $-$4.06 & 0.65 \\
       &        &           & $1/33$    & &   444 & $-$4.51 & 0.68 \\
       &        & 0.665     & $1/100$   & &   464 & $-$4.84 & 0.64 \\
600    & 7.21   &           & $3$    & 618 &   531 & $-$2.98 & 0.69 \\
       &        & 0.667     & $1$       & &   453 & $-$3.37 & 0.60 \\
       &        &           & $1/3$     & &   302 & $-$3.63 & 0.61 \\
       &        & 0.679     & $1/10$    & &   518 & $-$4.16 & 0.76 \\
       &        &           & $1/33$    & &   449 & $-$4.45 & 0.67 \\
       &        & 0.681     & $1/100$   &  &  482 & $-$4.77 & 0.67 \\
750    & 7.34   &           & $3$        & 645 &   337 & $-$2.65 & 0.70 \\
       &        & 0.706     & $1$        & &   287 & $-$2.93 & 0.65 \\
       &        &           & $1/3$      & &   326 & $-$3.48 & 0.62 \\
       &        & 0.718     & $1/10$     & &   214 & $-$3.78 & 0.84 \\
       &        &           & $1/33$     & &   434 & $-$4.25 & 0.71 \\
       &        & 0.719     & $1/100$    & &   435 & $-$4.58 & 0.67 \\
800    & 7.37   &          & $3$        & 653 &   502 & $-$2.82 & 0.68 \\
       &        & 0.715    & $1$        & &   315 & $-$3.02 & 0.79 \\
       &        &          & $1/3$      & &   351 & $-$3.34 & 0.64 \\
       &        & 0.728    & $1/10$     & &   402 & $-$3.88 & 0.62 \\
       &        &          & $1/33$     & &   488 & $-$4.26 & 0.67 \\
       &        & 0.729    & $1/100$    & &   456 & $-$4.56 & 0.69 \\
900    & 7.43   &          & $3$        & 669 &   524 & $-$2.78 & 0.66 \\
       &        & 0.732    & $1$        & &   377 & $-$2.95 & 0.65 \\
       &        &          & $1/3$      & &   284 & $-$3.19 & 0.62 \\
       &        & 0.745    & $1/10$     & &   237 & $-$3.65 & 0.70 \\
       &        &          & $1/33$     & &   453 & $-$4.11 & 0.70 \\
       &        & 0.746    & $1/100$    & &   457 & $-$4.50 & 0.70 \\
\hline
94     & 6.11   & 0.337    & $1/10$     & &   496 & $-$5.67 & 0.63 \\
       &        & 0.338    & $1/100$    & &   591 & $-$6.41 & 0.89 \\
193    & 6.56   & 0.463    & $1$        & &   610 & $-$4.18 & 0.96 \\
489    & 7.11   & 0.648    & $1$        & &   298 & $-$3.19 & 0.63 \\
588    & 7.21   & 0.681    & $1$        & &   406 & $-$3.35 & 0.60 \\
737    & 7.34   & 0.706    & $1$        & &   324 & $-$2.91 & 0.69 \\
\hline
\end{tabular}
\caption{Mass-loss predictions for VMS with parameters from the Gr\"afener et al. (2011) $M$-$L$ relationship.
A few extra models are shown below the line at the bottom of the Table.  
\teff\ is kept constant at 15,000 K.}
\label{tab:results}
\end{table*}

\begin{figure}
\centerline{\psfig{file=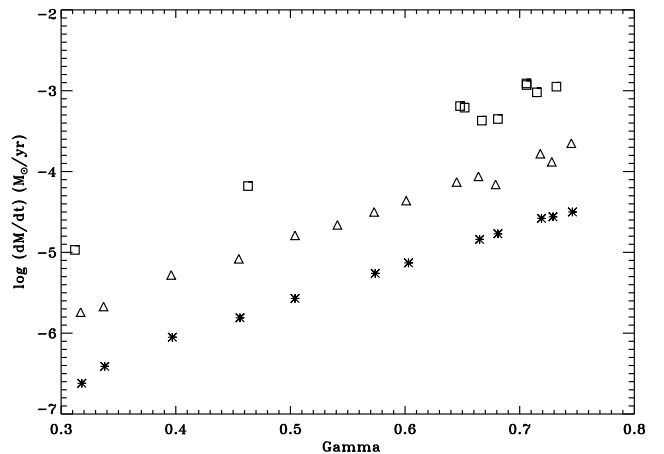, width = 9 cm}}
\caption{Predicted mass-loss rates ($\mdot_{\rm wind}$) versus $\Gamma$ for solar metallicity models (open squares), 
models of 10\% \zsun\ (open triangles) and 1\% \zsun\ (asterisks). The mass-loss rates show the expected linear drop
with lower $Z$.}
\label{f_mdot}
\end{figure}

\begin{figure}
\centerline{\psfig{file=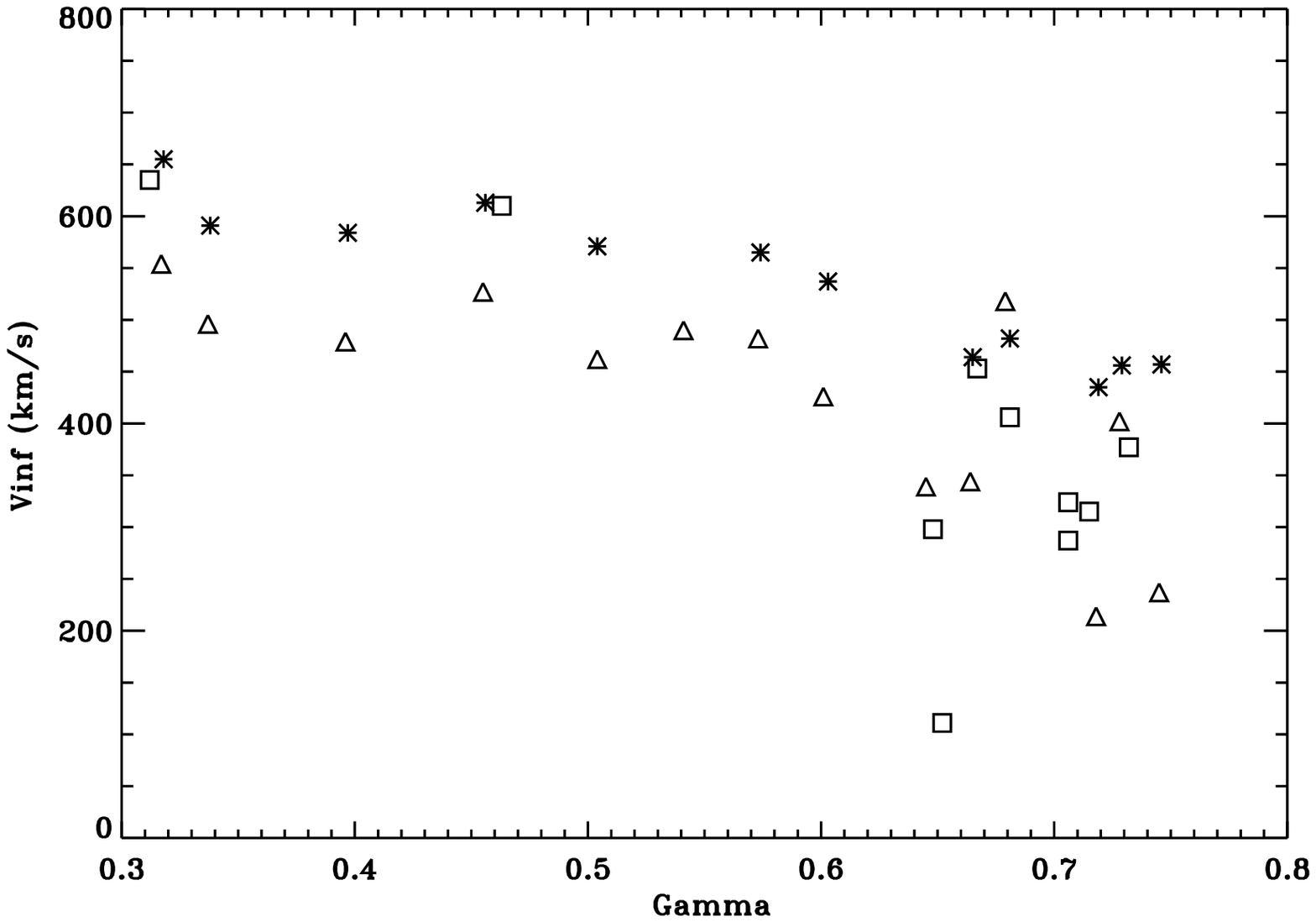, width = 9 cm}}
\caption{Predicted wind velocity ($\vinf$) versus $\Gamma$ for solar metallicity models (open squares), 
models of 10\% \zsun\ (open triangles) and 1\% \zsun\ (asterisks). We note that the wind velocities do not show a linear 
drop with $Z$. The reason is explained in the main body text.}
\label{f_vinf}
\end{figure}

Table~\ref{tab:results} lists our mass-loss predictions. 
The initial stellar parameters are given in columns (1) - (5).
The predicted wind terminal velocities, mass-loss rates, and 
wind acceleration parameter $\beta$ are 
listed in columns (6), (7), and (8). 
The predicted mass-loss rates are shown in Fig.~\ref{f_mdot}, and 
the resulting terminal wind velocities are shown in Fig.~\ref{f_vinf}.
Different symbols are used to identify the different $Z$ ranges.
For clarity, not all $Z$ models feature here, but only alternating 
$Z$ values from Table~1 have been plotted.

Figure 1 shows that $\mdot$ increases with $\Gamma$ and $Z$, as expected (Abbott 1982; Vink et al. 2001; Kudritzki 2002). 
Generally, the predicted mass-loss rates are in good agreement with the non-dynamically consistent
metallicity-dependent mass-loss rates of Vink et al. (2000; 2001). We note however that the 2000/2001 recipe
that was provided alongside the predicted rates in the form of an IDL routine bases its
output on the location of the second bi-stability jump (Vink et al. 1999; Petrov et al. 2016). This means that 
where the 2000/2001 recipe
indicates a location below the second bi-stability jump, it gives larger values (by a factor of a few) 
than predicted for a fixed value of 15\,000 K. It is thus important to assess the temperature location
of bi-stability jumps before for example comparing them to observed values. This also shows that 
the true mass-loss rates of bloated VMS may be even higher than predicted in this paper (if the 
stars make it to cooler temperatures than considered here).
Moreover, we note that at these relatively cool temperatures our models remain optically thin due to the large
stellar radii, and kinks in the $\dot{M}$ versus $\Gamma$ relation which were present for 
hotter VMS models (Vink et al. 2011), are thus absent. 

We now turn our attention to the wind velocity structure parameter, $\beta$, which 
describes how rapidly the wind accelerates. 
The predicted values of $\beta$ are given in column (8) of the Table. 
$\beta$ values are mostly in the range 0.6-0.8, in accordance 
with the models of Pauldrach et al. (1986), M\"uller \& Vink (2008) and Muijres et al. (2012). 

Figure 2 shows relatively low terminal wind velocities in the range 400-600 km/s for moderately 
high (0.3 - 0.6) Eddington models, going down to 100-200 km/s at even larger 
Eddington values. The main reason for these 
low values has little to do with metallicity (see below). The main reasons are (i) the low effective gravity \& 
escape velocity for these large stars, and the (ii) high mass-loss rates for these high $\Gamma$ objects instead.

These slow winds are most relevant
for a potential role of VMS for enriching Globular Clusters, as fast winds from normal O-stars
might escape the potential well of the young Globular Cluster. These slow winds might also be relevant
for the narrow He {\sc ii} line emission seen at intermediate (Cassata et al. 2013) and high (Sobral et al. 2015) redshift
as discussed in Gr\"afener \& Vink 2015). 

Interestingly the wind terminal velocities do not display a linear drop with $Z$, as is the case 
for $\dot{M}$, shown in Fig.~1. Figure~2 shows that the triangles representing intermediate values 
of 10\% $\zsun$ to be lower than both the high (solar) $Z$ squares, as well as the very low 1\% $\zsun$
asterisks. 
The reason
for this non-linear behaviour is that there are two competing physical effects. The first one is 
the direct line driving effect: less efficient driving at low $Z$ leads to smaller terminal velocities.
The second effect is that due to the lower $\dot{M}$ at lower $Z$, the lower 
density at the critical (sonic) point leads to a larger terminal wind velocity, as the driving of optically thick lines -- 
relevant for the supersonic portion of the wind that determines the terminal wind velocity -- is inversely proportional to the
density. This means there must exist a 
minimum wind velocity for our objects.
In our set of model calculations, the minimum wind velocities are reached at $Z/\zsun$ of 
$\simeq$ 1/33, that is, at globular cluster metallicities of order [Fe/H] $\simeq$ $-$1.5. 

\subsection{Mass loss recipe for VMS as a function of metallicity}
\label{s_recipe}

In order to determine the dependence of the mass-loss rate on $M$ and $Z$ simultaneously,
we perform multiple linear regression, finding
 
\begin{equation}
\label{eq:fit}
\log \mdot~=~-9.13~+~2.1 \log(M/\Msun)~+~0.74 \log(Z/\zsun)~~[\msunyr]  
,\end{equation}
with a fitting error of $\sigma$ = 0.09. The formula was derived for the $Z$ 
range $(Z/\zsun) = 1 - 10^{-2}$ and the mass range 100 - 900\,\msun. Extrapolation into the 
regime of SMS above $10^3$ \msun\ range is at potential users' own risk.
   
The mass-loss versus luminosity or Eddington $\Gamma$ relationship  
can be transformed using relevant mass-luminosity relationships (Gr\"afener et al. 2011).

\subsection{Comparison to observations \& previous models}

It is not possible to directly compare our new predictions against observations
or previous model predictions as models in this parameter range have not
been calculated before. The mass-loss rates predicted here for 15 kK are larger than
those computed by Vink et al. (2011), with the main reason being
the bi-stability jump (Vink et al. 1999). Terminal wind velocities in LBVs undergoing
S Doradus type variations are generally of the order of 100-500 km/s (see Vink 2012), lending 
empirical support for our wind predictions.

Gr\"afener \& Hamann (2008) 
computed optically thick wind models for hotter WNh stars as a function of $Z$, finding them 
to be lower than those of Vink et al. (2000, 2001). Also the optically thin wind 
models by Pauldrach et al. (2012) gave lower mass-loss rates than Vink et al. (2000, 2001)
for hot VMS in the hotter (40-50 kK) range.

\section{The 'effective' upper mass limit of stars}
\label{s_upper}

For nucleo-synthesis of VMS (Woosley \& Heger 2015) and the associated (maximum) metallicity-dependent 
mass limits of black holes (Eldridge \& Vink 2006; Belczynski et al. 2010),
the important parameters are the maximum mass that star formation allows for as well as the 
subsequent mass-loss history.
Whether the most massive stars form by core accretion, mergers, or coalescing stars in 
dense clusters, we argue that the most meaningful parameter is the 
effective upper-mass limit for stars with different $Z$, that incorporates both star formation and 
early stellar evolution including stellar wind effects.

This effective upper-mass limit is set not only by $Z$, but also by the absolute value of the 
mass-dependent mass-loss rate, derived above. 
This radiation-driven mass-loss rate may already be relevant during the PMS 
evolution during massive star formation, or otherwise during the core-hydrogen
main sequence evolution of merger products. 
The final result will depend on the exact physical mechanism
that sets the mass-accretion rate during star formation, whether or not it is 
$Z$ dependent (de Marchi et al. 2011, but see Kalari \& Vink 2015). 

However, whilst it not yet established if the mass-accretion rate is $Z$ dependent, we may 
at least conclude that the effective upper-mass limit is $Z$ dependent, with 
a higher effective upper-mass limit for lower $Z$ galaxies.
The quantitative value of the upper-mass limit -- for each $Z$ -- will however depend on the mass-dependent
mass-loss rate. 

\begin{figure}
\centerline{\psfig{file=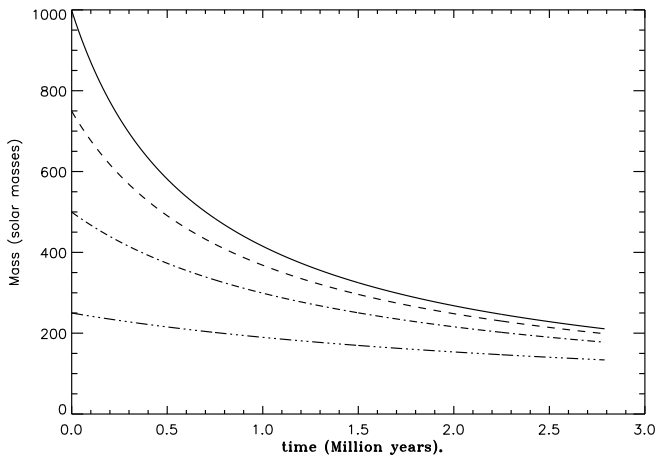, width = 9 cm}}
\caption{`Effective' upper-mass limit at solar metallicity. 
The effective limit drops with time, 
as four different initial masses (250, 500, 750, 1000 \msun) -- represented by the 
four different lines -- `converge' to very similar 
values. The figure also highlights that it is not possible to infer the value of the initial 
stellar upper-mass limit from
present-day observations (unless one has the exact theoretical knowledge of the mass-loss history available).}
\label{f_evap}
\end{figure}

During stellar evolution (and/or formation) the stellar mass itself is dropping, thereby continuously lowering its 
mass-loss rate. In other words, when we account for the time evolution
of a star (basically integrating the mass-loss rate over time), one observes a certain 
`softening' or `convergence' of the drop in mass 
with time for various initial masses, as depicted in Fig.\,\ref{f_evap}. 
Independent of the initial mass, the mass after 2 Myrs is 
of order 200 \msun. This behaviour is similar to that found in Yungelson et al. (2008), but their mass-loss prescriptions
were entirely ad-hoc, whilst Fig.\,3 of the current study shows an (arguably still simplified) 
evolution that is based on actual mass-loss computations, albeit at constant effective temperature.

Figure 3 also shows that from current VMS observations (e.g. Schneider et al. 2018) it is not possible
to infer the quantitative value of the upper-mass limit. 
In fact, it is not possible to determine this value quantitatively,
until we know both the mass-loss history\footnote{The mass-loss history will depend on the combined effects of VMS structure 
and evolution as well as the mass-loss rate.}, as well as the mass-accretion rate during star formation.

An important aspect will be to properly investigate if VMS spend enough time at these 
cool temperatures to have a significant quantitative effect on the upper-mass limit. At the moment we do not
yet understand massive star structure \& evolution sufficiently well to be confident in our current evolutionary models.
Not only is massive star evolution strongly dependent on $Z$, but the basic question of whether inflated
envelopes are stable, or perhaps removed by mass loss, is still unresolved. 
Ultimately, final answers will depend
on the physics of stellar envelopes and winds, both individually, as well as via their combined effect.
On the positive side, what we can conclude already is that -- independent of whether 
the mass-accretion rate is $Z$-dependent -- the upper-mass limit is expected to be 
$Z$-dependent due to the intrinsic nature of line-opacity mediated radiation-driven winds.

\section{VMS self-enrichment of globular clusters}
\label{s_self}

Over the past decade, two of the main contenders for the self-enrichment of globular clusters
have been massive AGB stars (e.g. d'Ercole et al. 2010) 
and as an alternative some form of `massive stars'. One of the 
main attractions of AGB stars over massive stars was that AGB stars have slow winds, whilst 
the line-driven winds of massive O-type stars are fast, up to 2000-3000 km/s. They are so fast, that 
the polluted wind material cannot be kept within the potential wells of either currently observed or young globular 
clusters\footnote{In addition to these intrinsically slower winds, 
the winds may shock and collide, leading to slow outflows that may be retained in the potential wells of globular 
clusters, allowing for a second
epoch of star formation (W\"unsch et al. 2017; Lochhaas et al. 2018).}. 
This is probably one of the main reasons why alternative massive star scenarios, such as the 
rapidly rotating massive stars (Decressin et al. 2007), massive binaries (de Mink et al. 2009), and 
red supergiants (Sz\'ecsi et al. 2018), have been considered. All these scenarios of course have there 
pros and cons and for an extensive overview see Bastian \& Lardo (2018) and Gratton et al. (2004). 

One of the main problems for all these scenarios 
is the so-called mass-budget problem and Gieles et al. (2018) recently proposed that a good way around 
it is the presence of a supermassive star (SMS) working as a `conveyer belt' continuously accreting fresh material \& polluting 
the cluster with enriched material. 
Although this is a very attractive scenario, the estimated wind velocities of 1000 km/s are too large in comparison to 
their estimated escape velocities at the centres of young globular clusters of order 500 km/s for a cluster mass of $10^6$ \msun\ 
(Gieles et al. 2018). Moreover, 
SMS are still a hypothetical concept whilst VMS, at least
up to 200-300\msun\ {\it do} exist (Crowther et al. 2010; Bestenlehner et al. 2011; Oskinova et al. 2013; Vink
et al. 2015; Martins 2015; Crowther et al. 2016). 
Furthermore, there appear to be more of them than expected from a Salpeter IMF 
(Schneider et al. 2018) with enhanced mass-loss rates, probably enabling us to overcome the mass budget problem (see Introduction).

It thus appears to be appropriate to consider VMS as the main culprit polluting
globular clusters as they dominate the wind feedback (Doran et al. 2013). The structure of VMS involves 
the inflation of the outer envelope due to the Fe opacity and the proximity to the Eddington limit 
(Ishii et al. 1999; Sanyal et al. 2015; Gr\"afener et al. 2012). This implies stellar effective temperatures 
lower than the usual 40-50 kK  ZAMS as discussed in the current paper. 
These cooler massive B supergiants have slower winds, from which the wind material may be 
expected to be maintained in the potential wells of globular clusters. 
B supergiants as discussed here have the advantage over red supergiants (RSGs) as recently discussed by 
Sz\'ecsi et al. (2018)
of being present in the observable universe, for example, as LBVs (Humphreys \& Davidson 1994), whilst 
RSGs are not known to exist above the Humphreys-Davidson limit. 

Although the details of our proposed scenario need to be worked out in terms of self-consistent
VMS structure \& evolutionary models, the fact that 
they must have played a role seems hard to deny, given that they almost certainly existed, whilst 
alternative scenarios involving either very rapidly rotating stars, RSGs, and possibly even SMS may remain largely hypothetical. 
The existence of massive binaries in globular clusters is also not yet established, but appears less exotic.

Finally, one of the main reasons for the popularity of the fast-rotating massive stars scenario of Decressin et al. 
(2007) was that it combined two aspects of rapid rotation: (i) it could transport the H-burning core material to the outer layers, whilst 
rotationally supported disk winds would pollute the globular cluster at low speeds. 
The issue is that rapid rotation seems empirically rather rare, especially at relatively high $Z$
(see Ram\'irez-Agudelo et al. 2013) 
By contrast, VMS form the natural extension of massive stars, and as we have shown here, they are expected to have relatively 
slow winds. 

Finally, rotation-induced mixing may no longer be required for the highest masses, as VMS have 
both large convective cores \& vigorous mass loss. 
This ensures that their evolution is close to chemically homogeneous 
(Gr\"afener et al. 2011; Yusof et al. 2013; Hirschi 2015; Vink 2015),  
regardless of the stellar rotation rate (Vink \& Harries 2017).  

\section{Summary}
\label{s_sum}

We present mass-loss predictions from Monte Carlo radiative 
transfer models for cool VMS in the $10^2-10^3$ \msun\ range, and we find that:

\begin{itemize}

\item{The mass-loss rate is expressed as a function of mass and $Z$ through 
multiple linear regression}.
 
\item{We find $\log$ \mdot\ $=$ $-$9.13 $+$ 2.1 $\log(M/\Msun)$ $+$ 0.74 $\log(Z/\zsun)$,        
derived for $(Z/\zsun)$ $=$ $1 - 10^{-2}$ and the range 100 - 900\,\msun.}

\item{We predict mass-loss rates that rival mass-accretion rates of $10^{-3}$ $\msunyr$ during
 massive-star formation, with important consequences for the stellar upper-mass limit.}

\item{We predict relatively slow terminal wind velocities ($\vinf$) in the range 100-500 km/s}

\item{We propose slow winds from VMS as a source of chemical pollution of globular clusters.}

\end{itemize}


\begin{acknowledgements}

We would like to thank the anonymous referee for constructive questions and comments that helped improve the paper.

\end{acknowledgements}

\end{document}